\documentclass[reprint,amsmath,amssymb,aps,floatfix]{revtex4-2}

\usepackage{graphicx}% Include figure files
\usepackage{dcolumn}% Align table columns on decimal point
\usepackage{bm}% bold math
\usepackage{esvect}
\usepackage{amsmath}
\usepackage[mathscr]{euscript}
\usepackage[italicdiff]{physics}
\usepackage{mathtools}
\usepackage[colorlinks=true,linkcolor=blue,citecolor=blue]{hyperref}% add hypertext capabilities
\def\orcid#1{\kern .08em\href{https://orcid.org/#1}{\includegraphics[keepaspectratio,width=1em]{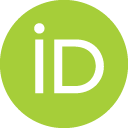}}} 

\begin{document}

\preprint{APS/123-QED}

\title{Driven particle in a one dimensional periodic potential with feedback control: 
efficiency and power optimization}
\author{Kiran V \orcid{0000-0003-3960-2589}}
\email{kiran.vktm@gmail.com, p20180029@goa.bits-pilani.ac.in}
\affiliation{Department of Physics, BITS Pilani K K Birla Goa Campus, Zuarinagar 403726, Goa, India}
\author{Toby Joseph \orcid{0000-0001-6682-9223}}
\email{toby@goa.bits-pilani.ac.in}
\affiliation{Department of Physics, BITS Pilani K K Birla Goa Campus, Zuarinagar 403726, Goa, India}

\date{\today}

\begin{abstract}
A Brownian particle moving in a staircase-like potential with feedback control offers 
a way to implement  Maxwell's demon. An experimental demonstration of such a 
system using sinusoidal periodic potential carried out by 
Toyabe et al. [\href{https://www.nature.com/articles/nphys1821}{Nature Physics 6, pages 988–992 (2010)}] 
has shown that information about the particle's position can be converted to useful work. In this paper, 
we carry out a numerical study of a similar system using Brownian dynamics simulation. 
A Brownian particle moving in a periodic potential under the action of a constant 
driving force is made to move against the drive by measuring the position of the particle 
and effecting feedback control by altering potential. The work is extracted during 
the potential change and from the movement of the particle against the external drive. 
These work extractions come at the cost of information gathered during the measurement. 
Efficiency and work extracted per cycle of this information engine are optimized by 
varying control parameters as well as feedback protocols. Both these quantities are 
found to crucially depend on the amplitude of the periodic potential as well as the 
width of the region over which the particle is searched for during the measurement 
phase. For the case when potential flip ({\it i.e.}, changing the phase of the 
potential by $180$ degrees) is used as the feedback mechanism, we argue that the 
square potential offers a more efficient information-to-work conversion. 
The control over the numerical parameters and averaging over large number of trial 
runs allow one to study the non-equilibrium work relations with feedback for this 
process with precision. It is seen that the generalized integral fluctuation theorem 
for error free measurements holds to within the accuracy of the simulation.

%\begin{description}
%\item[Usage]
%Secondary publications and information retrieval purposes.
%\item[Structure]
%You may use the \texttt{description} environment to structure your abstract;
%use the optional argument of the \verb+\item+ command to give the category of each %item. 
%\end{description}
\end{abstract}

\maketitle

%\tableofcontents

\section{Introduction}
The feedback control associated with Maxwell's demon like setups allows one to extract
heat from a thermal bath and convert it into useful work \cite{ClerkMaxwell:1871:TH,Maruyama2009}. 
In the Szilard engine version of the Maxwell demon implementation, the demon determines 
whether a single molecule present in a vessel which is in contact with the thermal bath 
is on the left or right half and uses that information to extract work via isothermal 
expansion of one of the pistons at the ends of the vessel \cite{szilard1964decrease}. 
The engine apparently seems to violate the second law of thermodynamics. After almost half 
a century of controversies and discussions, the paradox has been resolved with the understanding
(for alternative views, see the following references \cite{earman1998exorcist,earman1999exorcist,
hemmo2010maxwell,norton2011waiting,kish2012energy}) 
that it is possible to extract work from such a system without contradicting the second law of 
thermodynamics, provided one has accounted for the cost of information processing carried out 
by the demon \cite{Landauer1961,bennett1982thermodynamics,Maruyama2009,sagawa2009minimal,mmaxwell}. 
In another words, it is possible to convert information into free energy or extract 
work using the available information. Though the problem itself is more than a century old, 
experiments implementing the demon were only achieved fairly recently both in classical 
\cite{toyabe2010experimental,berut2012experimental, saha2021maximizing,paneru2018lossless} and 
quantum systems \cite{koski2014_zilard_experimental,PhysRevResearch.2.032025,averin2011maxwell,
camati2016experimental,naghiloo2018information, chida2017power}.  

One of the first experimental studies of a classical system that converts information into free 
energy using feedback control was demonstrated by Toyabe et al. \cite{toyabe2010experimental}. 
In this experiment, a colloidal particle in contact with a thermal bath undergoes rotational 
Brownian motion in a stair-case like potential with the step height comparable to $k_{B}T$.
The stair-case like potential is created by a combination of a sinusoidal potential and a linear one. 
The particle can take energy from the thermal bath and make an upward jump or can slide down 
in the direction of the negative gradient of the potential. In the experiment, one selectively 
manipulates such fluctuations via position measurement of the particle and subsequent 
feedback control to extract work from the heat bath. The feedback control was carried out by 
changing the phase of the sinusoidal potential by $180$ degrees (referred to as potential flip), 
depending on the outcome of the measurement of particle's angular position. The feedback control 
helps to extract useful work from the thermal bath via two routes: (i) The work done against 
the linear potential (which the authors refer to as the free energy gain, $\Delta F$) and 
(ii) as work extracted during potential flip (referred to as $-W$). The work extracted 
is accounted for by the  energy equivalent of information obtained during the measurement 
process and there is no violation of the second law. The efficiency of conversion of information 
to work extracted of the engine is $28\%$. One of the motivations for the present work
is to use simulations to better understand the low efficiency values and explore ways on 
optimizing this engine.

Many recent works, both in experiments \cite{saha2021maximizing,paneru2018optimal,rico2021dissipation} 
and in theory \cite{lucero2021maximal,dinis2020extracting,pal2014extracting}, have looked at 
ways to improve efficiency and power of information engines. In the domain of classical information 
engines, the Brownian information engine based on a colloidal particle in a harmonic potential 
has been studied extensively \cite{abreu2011extracting,bauer2012efficiency}. But similar detailed 
study on optimisation of the information engine based on particle moving in a periodic potential 
is lacking. One way to improve the low value of efficiency is by fine tuning the parameters and 
optimizing the control protocol of the feedback processes. A general feedback scheme for extracting 
maximum work is by changing the Hamiltonian of the system right after the measurement in such a way 
that the post measurement state is an equilibrium state of the new Hamiltonian 
\cite{Parrondo2015,abreu2011extracting, hasegawa2010generalization,esposito2011second,horowitz2011designing}. 
Such a protocol is completed by reversibly adjusting the external parameters to the final values. 
In the present work, we vary the shape of the periodic potential as well as the parameters in the 
feedback protocol to achieve optimal conversion of information to work based on this principle.

Advances in the area of stochastic thermodynamics in the last few decades have
augmented our understanding of how irreversibility emerges from reversible dynamics 
\cite{jarzynski2011equalities,Seifert2012}. Various fluctuation theorems have provided insights 
about entropy production and statistical relationships between work and free energy for systems 
driven far away from the equilibrium 
\cite{evans1993probability,gallavotti1995dynamical,Jarzynski1997,crooks1999entropy}. 
The Jarzynski equality (JE) given by $\left<e^{(\Delta F - W)/k_{B}T} \right> = 1$ was one of the first 
work relations to be derived and it relates the fluctuations in work during a non-equilibrium process 
to the free energy difference between the final and initial equilibrium states \cite{Jarzynski1997}. 
This relation breaks down in the presence of feedback process. For processes involving error 
free measurement and feedback, one can derive a generalized integral fluctuation theorem (GIFT)
given by $\left<e^{(\Delta F - W)/k_{B}T - I + I_u}\right> = 1$, were $I$ is the information 
gained during the measurement process and $I_u$ is the unavailable information measured using 
the time reversed process (see discussion in Sec.\ref{vgift} for details) \cite{ashida2014general}. 
The JE itself takes a modified form given by  $\left<e^{(\Delta F - W)/k_{B}T} \right> = \gamma$, 
where $\gamma$ (referred to as efficacy) measures the  reversibility of the  process. 
Experimental verification of these relations have been done for a few systems 
\cite{toyabe2010experimental,Koski2014,Paneru2020,paneru2020colloidal}. The simulations 
presented here allow for checking the validity of these generalized fluctuation theorems
with precision for the studied system.

The paper is organized as follows. The following section introduces the model. 
In Sec. \ref{Results}, we discuss the details of the simulation and the results. Drift of 
the particle per cycle and efficiency of information to work conversion for different 
values of feedback delay are studied. Various optimization studies of efficiency are 
presented in Sec. \ref{optimization}. These include improvement of efficiency by optimizing the parameters 
of the model and alteration of feedback protocol. The section ends with the discussion of 
efficiency of a similar system but with a square potential. In Sec. \ref{vgift}, the  generalized 
integral fluctuation relation is discussed and verified for this system 
and in Sec. \ref{conclusion} we summarize and discuss the results.

\section{The Model}\label{model}
\begin{figure}
\includegraphics[scale=0.4]{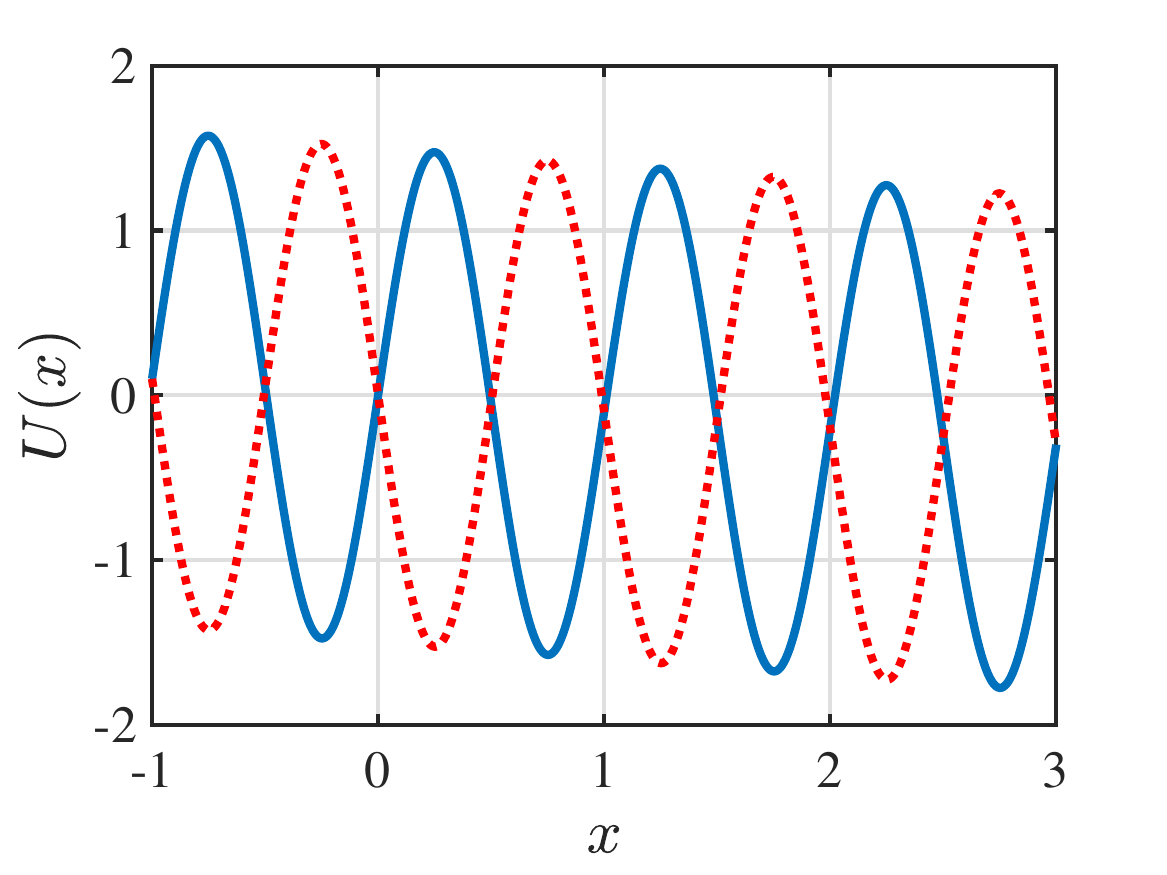}
\caption{\label{fig:potential} Tilted sinusoidal potential with period equal to $1$ unit. 
The amplitude of the sinusoidal part and slope corresponding to the uniform force 
are $1.5$ (in units of $k_B T$) and $0.1$ (in units of $k_B T$ per period of the potential) 
respectively. The red and blue curves correspond to the potential before and after 
switching of the phase of the sinusoidal part of the potential.}
\end{figure}

Consider a particle moving in a sinusoidal potential under the influence of a 
constant driving force. The net potential in which the particle moves is given by 
(see Fig. (\ref{fig:potential}))
\begin{equation}
U(x) = \pm U_0 \sin{(2\pi x}) - F_d x,
\label{eq:1}
\end{equation}
where $x$ is the position of the particle, $2U_0$ gives the depth of 
the periodic potential and $F_d$ is the magnitude of the driving force. The $\pm$ sign 
in the first term in the RHS is present because the phase of the potential is 
changed during the feedback process (see below). Additionally, the particle is in 
contact with a thermal bath kept at temperature $T$. 
The Langevin equation governing the motion of the particle is given by,
\begin{equation}
\displaystyle m\ddot{x}(t) = -m\xi \dot{x}(t) \pm 2\pi U_0 \cos{(2 \pi x(t))} + F_d + \zeta(t),
\label{laneq}
\end{equation}
where $m$ is the mass of the particle and $-m\xi \dot{x}$ is the viscous force. 
$\zeta(t)$ is the thermal noise with zero average and the correlation function is given by 
$\left<\zeta(t) \zeta(t')\right> = \Gamma \delta(t - t')$. Fluctuation-dissipation relation
connects the strength of the noise, $\Gamma$, to the friction coefficient, $\xi$, by the 
relation: $\Gamma = 2 m \xi k_{B}T$. In the over-damped limit, one can ignore the 
inertial term in Eq. (\ref{laneq}) and this leads to the Brownian dynamics (BD) equation, 
\begin{equation}
\dot{x} = \frac{\pm 2\pi U_0 \cos{(2 \pi x)} + F_d}{m\xi} + \frac{\zeta}{m\xi} \;.
\label{bdeq}
\end{equation}

\begin{figure}
\includegraphics[scale=0.4]{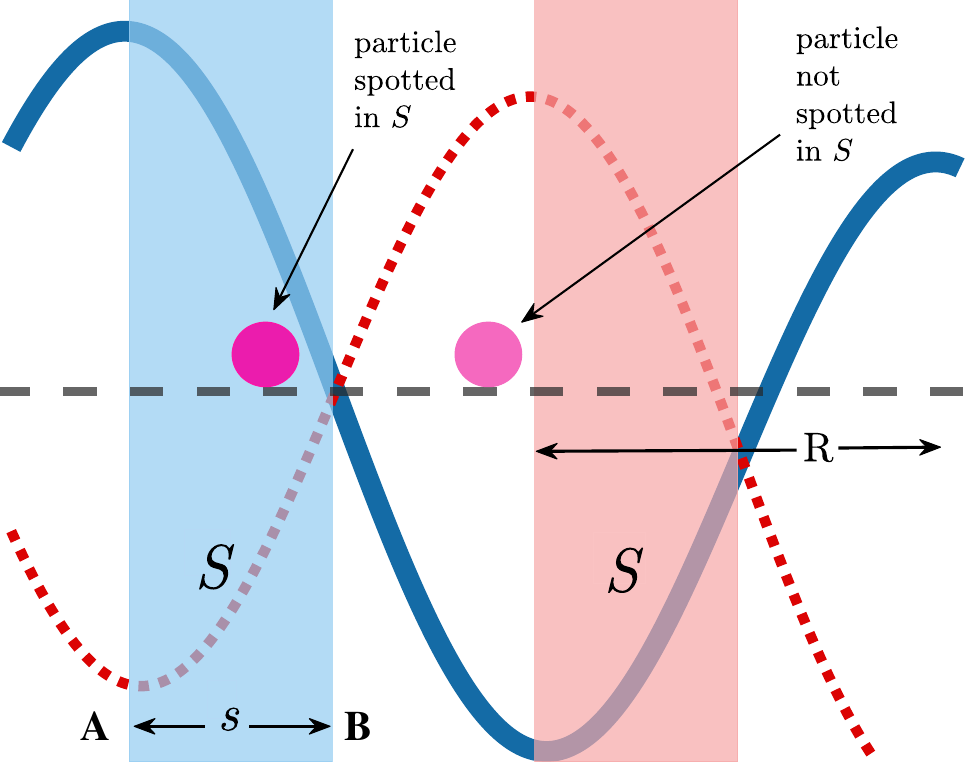}
\caption{\label{meas} For measurements carried out at time $t = n \tau$, the particle can 
be either spotted in region $S$ or outside of $S$. The feedback protocol is initiated according 
to the outcome of the measurement. The sinusoidal potential is flipped with a time delay of $\epsilon$, if the particle is spotted in $S$. The solid blue and dashed red curves correspond to the potential before and after the flip. The $S$ regions are indicated for both the potential configurations before and after flip, with the colours respect to the sinusoidal curves. $s$ is the width of the region $S$.
$R$ is the region to the right of the potential minimum.}
\end{figure}

The feedback process that is designed to help the particle move in the direction opposite 
to that of the externally applied driving force, $F_d \hat x$, and gain free energy in 
the process is as follows: At times given by $t = n \tau$, a measurement of particle's position 
is carried out. If the particle is located in the region $S$ (see Fig. (\ref{meas})), 
then the phase of the potential is changed by $\pi$ instantaneously (henceforth called 
potential flip) at a time $t = n\tau + \epsilon$, where $\epsilon$ is the feedback delay time. 
If the particle is not spotted in the region $S$, no feedback process is initiated. We shall later 
alter this feedback procedure in order to improve information to free energy conversion efficiency. 
These alterations would involve, in addition to the potential flips, the raising of the potential 
barrier when the particle is not spotted in $S$.

\section{Results from the simulation}
\label{Results}
The over-damped equation of motion, Eq. (\ref{bdeq}), has been integrated numerically 
using the discretized version \cite{ermak},
\begin{equation}
x(t + \delta t) = x(t) + \frac{\pm 2\pi U_0 \cos{(2 \pi x)} + F_d}{m\xi} \delta t +  f_g, 
\label{ermak}
\end{equation}
where $\delta t$ is the time step and $f_g$ is a Gaussian distributed random variable with 
zero mean and variance equal to $\frac{2 k_B T}{m\xi} \delta t$. We work with a system of 
units defined by $\xi = 1$, $m = 1$ and $k_{B}T = 1$. Length scale in the problem is set 
by the period of the potential, which is $1$ and the time scale is $\xi^{-1}$, which is
also $1$. The integration time step of the simulation is taken to be 
$\delta t = 0.0001$ and has been checked for convergence by carrying out simulations 
at one order less than this value. 
For the results quoted in this section, the amplitude of the sinusoidal 
potential is taken to be $U_0 = 1.5$ and the magnitude of the driving force is $F_d = 0.1$.
All the feedback processes in our study involve a single cycle and the duration of 
the cycle is $\tau = 0.05$. Before each cycle starts, we ensure that a long enough 
equilibration run is carried out so that correlation effects do not affect the 
results. The region $S$ starts from the maxima of the potential (point {\bf A} in 
Fig. (\ref{meas})) and ends at the point where the force is maximum (point {\bf B} in
Fig. (\ref{meas})), encompassing a total length of $s = 0.25$. 
The feedback delay time, $\epsilon$ is varied from the minimum value possible of $0.0001$ 
(since $\delta t = 0.0001$) to $0.047$. The mean values of various quantities of interest 
are determined by averaging over $10^6$ cycles.

\subsection{Particle drift and efficiency}
The physical quantities of relevance to study the stochastic thermodynamics of the system 
are: (i) the drift velocity of the particle (which is related to the rate at which the 
system gains free energy), (ii) the work done by the external agent in flipping the potential 
and (iii) the information gained during the measurement. The average drift velocity of the 
particle is given by 
\begin{equation}
v_d = \frac{\left< D \right>}{\tau},
\label{vd}
\end{equation}
where $D = x(\tau) - x(0)$ is the distance between the initial and final equilibrium locations 
of the particle. The angular brackets indicate average over the trials. Note that in the 
absence of the feedback process, the particle will drift in the direction of the drive and 
the average speed in the steady state can be exactly evaluated 
\cite{stratonovich1958synchronization,reimann2001giant}. But with feedback, the drift in the 
direction of drive can be reduced and even reversed, depending on how effective 
the feedback process is. The free energy gained per cycle, which is the work done against the
external force in one cycle, is given by
\begin{equation}
\Delta F = -F_d D,
\label{delf}
\end{equation}
which is positive if the particle has drifted against the direction of the drive.
The work done by the external agent is given by change in the potential energy of
the particle in the sinusoidal field at the instant the potential flip is carried out. 
That is,
\begin{equation}
W = \pm 2U_0 \sin{(2 \pi x(\epsilon))},
\label{wrk}
\end{equation}
where the $+$ sign is for the case when the potential after the flip is greater than its 
value before the flip, implying that the work is done by the external agent. If otherwise, 
the work is being extracted out of the heat bath. The work done in a given cycle is zero if 
the particle is not spotted in the region $S$ (as there is no potential flip carried out 
in this case).\\

The measurement carried out at the beginning of a cycle gives information about the location 
of the particle and is quantified by the Shannon information content, defined as $\left<I\right> 
= -p\ln{p} - (1 - p)\ln{(1 - p)}$ \cite{shannon1948mathematical} where $p$ is the probability for 
finding the particle in the region $S$. The energy equivalent of information content is 
$k_B T\left<I\right>$. For the values of $s$, $U_0$ and $F_d$ used in the results in this section,
the value of $k_B T \left<I\right> = 0.24$. In the measurement process, an amount of heat equal 
to $k_B T\left<I\right>$ is dissipated to the heat bath. This dissipation is associated 
with the erasure of memory bits required for the measurement. The feedback process allows 
one to regain a part of this dissipated energy back in the form of an increase in free energy 
of the particle and also possibly as work extracted from the heat bath. But this 
regain is possible provided the information obtained is used before the particle equilibrates 
after the measurement. 

\begin{figure}
\includegraphics[scale=0.4]{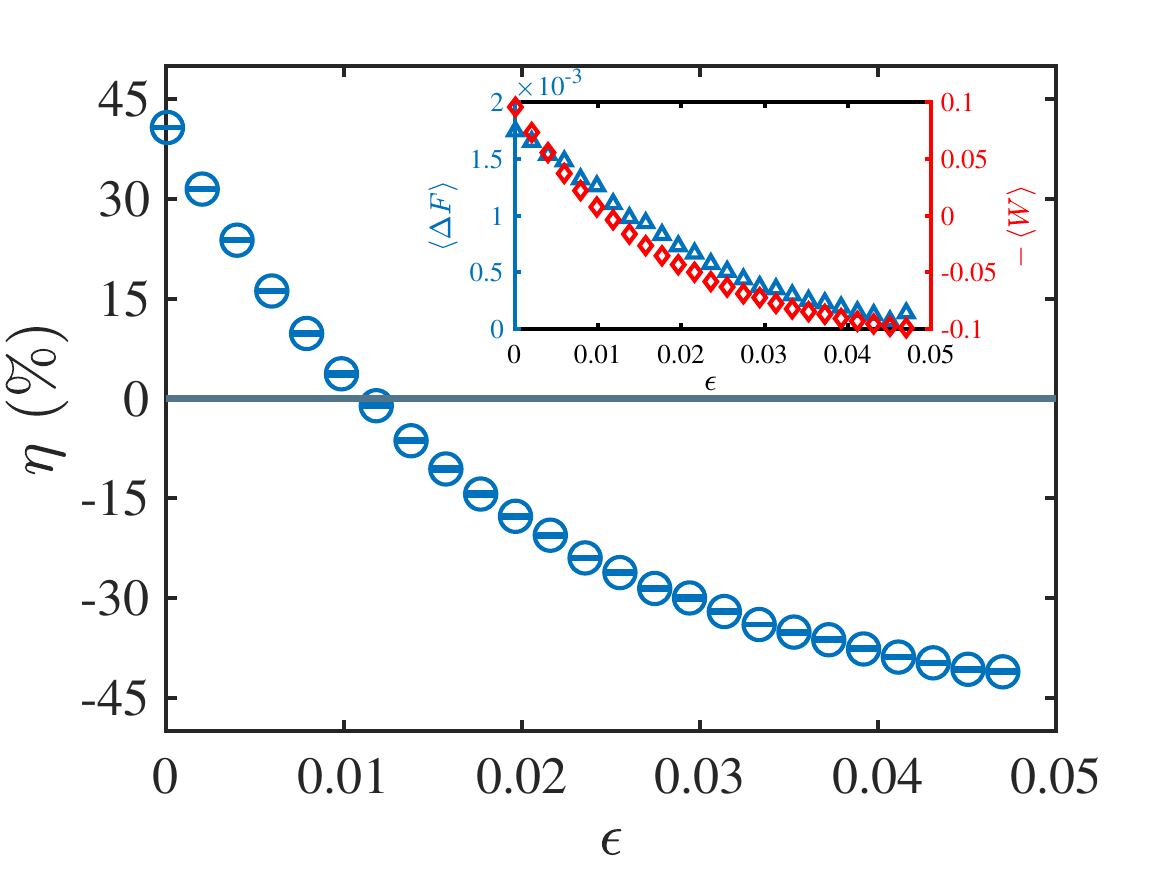}
\caption{\label{fig:ef1} Variation of efficiency with feedback delay time. For
$\epsilon \lessapprox 0.01$ (measured in units of $1/\xi$) the system works as an information 
engine converting information to work. For larger values of $\epsilon$, the efficiency 
becomes negative indicating that sum of gain in free energy and work extracted is negative.
The maximum value of efficiency obtained is $41 \%$ at $\epsilon = 0.0001$. The inset
shows the contributions of $\Delta F$ and $-\left<W\right>$ separately. For low values of $\epsilon$,  
$-\left<W\right>$ term is positive and one order larger than the $\Delta F$. For large values of $\epsilon$, 
$-\left<W\right>$ becomes negative implying that work has to be done by the external agent. The error bars
in the main figure are standard deviations.}
\end{figure}

\begin{figure}
\includegraphics[scale=0.4]{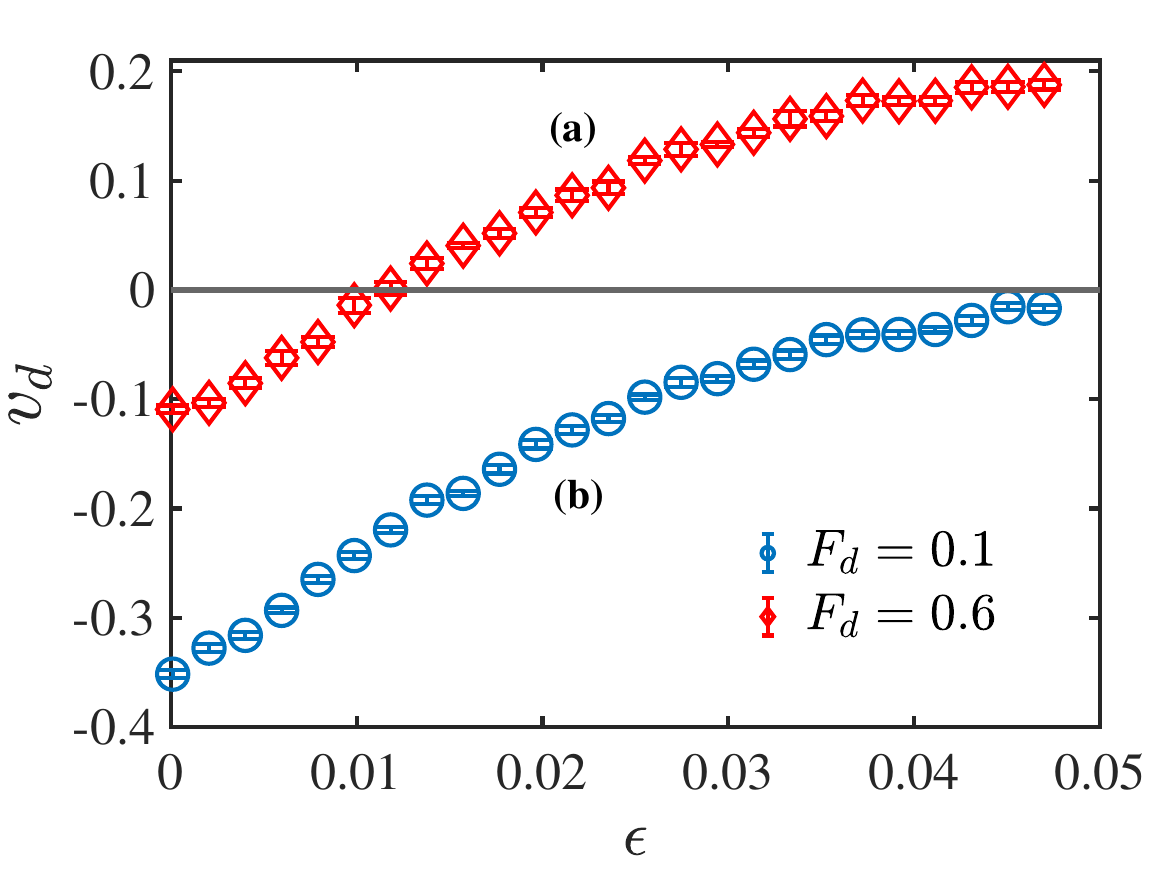}
\caption{\label{fig:velocity} Variation of particle's velocity (measured in units of period
of potential times $\xi$) as a function of feedback delay time, $\epsilon$. 
Negative value of displacement indicates net drift in the direction 
opposite to the uniform external force. For large values of $\epsilon$, the effect of feedback 
is reduced progressively. The $(a)$ and $(b)$ data points correspond to two different 
external drives, $F_d = 0.6$ and $F_d = 0.1$ respectively. The error bars in the figure are standard deviations.}
\end{figure}

The efficiency of the information engine can be defined as
\begin{equation}
\displaystyle\eta = \frac{\left<\Delta F - W\right>}{ k_{B}T\left<I\right>},
\label{etadef}
\end{equation}
which is a measure of how efficiently the available information is converted to free energy 
gain and work extracted. $\Delta F$ is the change in free energy and $-W$ is the work 
extracted, as defined above. Variation of efficiency and drift per cycle of the particle for 
different values of feedback delay $\epsilon$ are shown in 
Fig. (\ref{fig:ef1}) and Fig. (\ref{fig:velocity}) respectively. 
The efficiency is maximum for minimal feedback delay and maximum value of efficiency is 
around $41\%$ for a feedback delay of $0.0001$. The inset of Fig. (\ref{fig:ef1}) shows the individual
variation of $\Delta F$ and $-\left<W\right>$ with $\epsilon$. For low values of $\epsilon$, the 
contribution of $-W$ to the total work extracted is much greater than that of $\Delta F$. 
The system is not working as an information engine for large values of $\epsilon$ 
(the region where $\eta < 0$). In this regime $\left<\Delta F - W\right>$ is negative and 
heat is dissipated into the heat bath during the cycle. This is over and above the amount 
$k_B T \left<I\right>$ that needs to be dissipated for memory erasures associated with 
acquiring positional information during the measurement phase.

The monotonic decrease of the efficiency with increasing feedback delay can be 
understood as follows. Let us consider the situation that the particle is spotted in $S$ 
when the measurement is carried out at time $t = 0$. For small values of feedback delay, 
$\epsilon$, the particle has a significant chance to stay close to the current location 
(in $S$) by the time potential flip is carried out. As a result, the particle loses
potential energy during the potential flip. This means, work is done by the system 
($W$ is negative) (Refer Eq. (\ref{wrk})). Additionally, for small $\epsilon$ value, there 
will be a net drift towards the left on the average, as seen in Fig. (\ref{fig:velocity}) (circles). 
This is because, the particle will be spotted in $S$ only when it makes a leftward jump 
(against the applied force) with respect to its current equilibrium position and instantly 
switching the potential phase locally traps the particle in the new minima to the left. 
This helps the particle to move against the driving force direction which in turn results 
in a positive free energy change (since the displacement $D$ of the particle will be negative). 
Both of these reasons lead to a larger efficiency value for small feedback delays, for a 
given amount of information obtained. However for large values of $\epsilon$, the particle 
will most likely move to the right after it is spotted in $S$, because of the net force in 
that direction and will equilibrate in one of the minima of the potential. Consequently, 
the external agent has to do work ($W$ is positive) on the particle during potential flip, 
since the potential energy of the particle after the flip is likely to be larger). 
Also, the particle is more likely to move towards the right after the potential flip as 
there is a slight bias in the steady state distribution of the particle to the right 
due to the applied uniform force. This leads to a decrease in the free energy on the average. 
%It is interesting to note that the particle drifts more in the direction of force when 
%potential flips are carried out with large values of $\epsilon$ ($\epsilon \gtrapprox 0.1$) 
%compared to the case when feedback is \textit{not} implemented.

It is seen that even for low values feedback delay time, the efficiency of the information 
to free energy conversion is very less ($\approx 40 \%$). One reason for the low 
value of efficiency is the fact that there is no feedback implemented when the 
particle is not found in $S$ during the measurement. This means, one is not utilizing 
the full available information for feedback control. For the parameters used in above 
simulations, value of $p = 0.067$ and the corresponding value of $\left<I\right> = 0.24$. 
If we do not provide feedback when the particle is not found in $S$, the maximum information 
that we can hope to convert to free energy or work is $-p \ln{p} = 0.18$. This is about 
$75 \%$ of the total information gathered and so the unused information cannot fully 
account for the low value of efficiency seen. The primary reason for low efficiency is 
due to the fact that the feedback process is not reversible in the 
sense that the time reversed protocol do not lead to time reversed processes 
\cite{Parrondo2015} and thus the engine is working sub-optimally. One can modify 
the protocols to try and optimize the information to free energy conversion. 
We discuss these below.

\section{Optimization studies}\label{optimization}
In this section we address the following question: Given a fixed external drive 
($F_d = 0.1$, in our studies), how can one optimize the conversion of information to free energy? 
To start with, we fix the shape of the external potential to be sinusoidal and vary the parameters 
$s$ and $U_0$ sequentially to search for maximum efficiency. This achieves only a partial 
optimization as we are not scanning the entire parameter space. The intention is to see the scale 
of dependence of efficiency on these parameters. Next, we alter the feedback by associating a 
protocol when the particle is not spotted in the region $S$ during particle's position measurement, 
which enables more use of the available information for work extraction. We end the optimisation 
studies by finding efficiency of the system by using square potential instead of a sinusoidal 
shaped one. We shall argue that the square potential increases the reversibility of the feedback
process, leading to larger values of engine efficiency.

\subsection{Parameter optimisation for sinusoidal potential}
\begin{figure}
\includegraphics[scale=0.37]{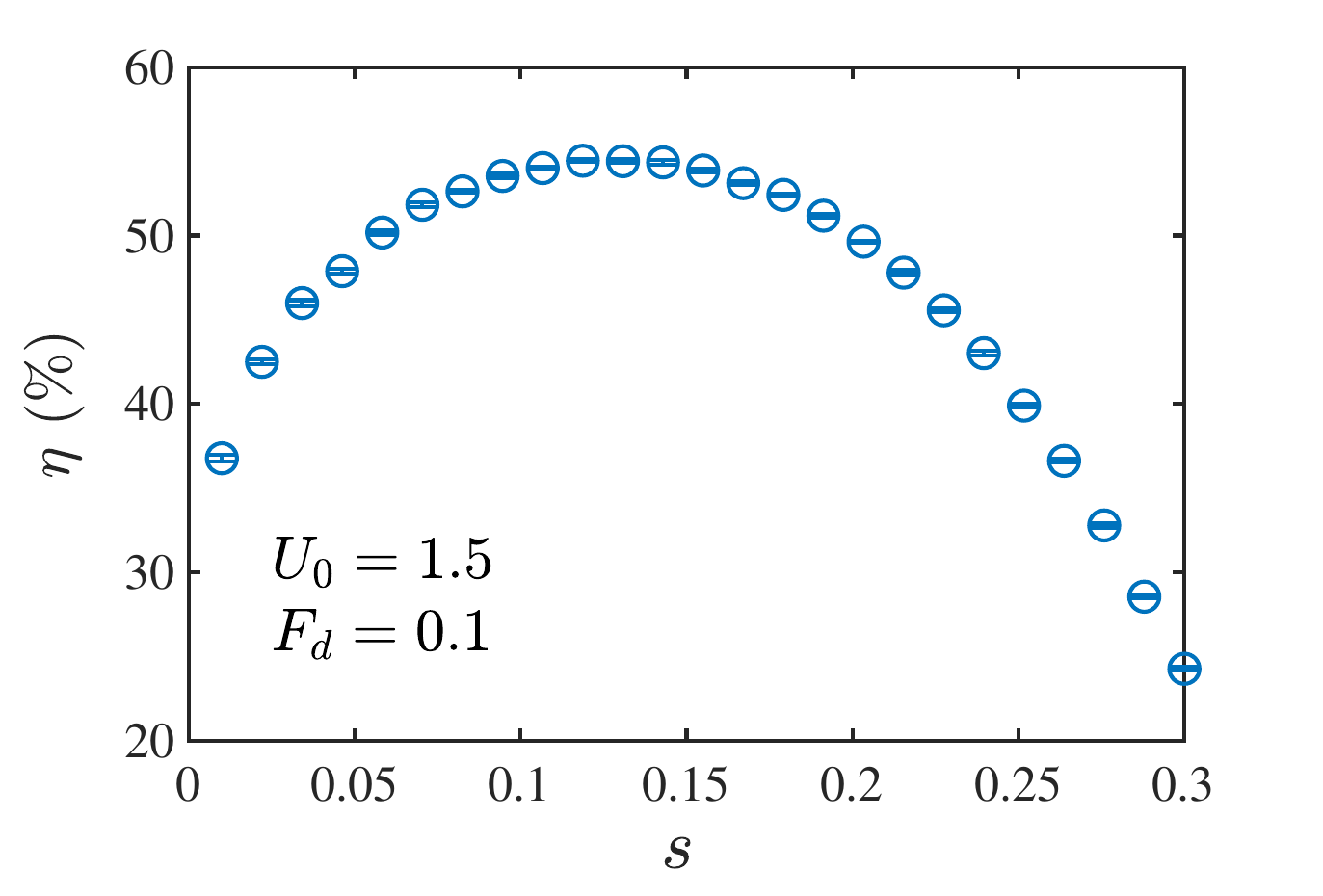}
\caption{\label{fig:optimization_s} Optimization of efficiency with respect to parameter $s$. 
For the values of $U_0 = 1.5$ and $F_d = 0.1$, a maximum efficiency value of $54.4\%$ is 
obtained for $s= 0.12$. The error bars in the figure are standard deviations.}
\end{figure}

The partial optimization with respect to experimental parameters was done as follows: 
First we find the optimum value of $s$ for which we get a maximum efficiency by keeping 
the values of $U_0$ and $F_d$ fixed at $1.5$ and $0.1$ respectively. $s$ is varied 
by keeping the location of the starting point of region $S$ fixed at the maxima of 
the potential. Efficiency is seen to have non-monotonic variation with $s$ and the 
maximum is found to be at $s = 0.12$ as shown in the Fig. (\ref{fig:optimization_s}).
The nature of variation of efficiency with $s$ can be understood as follows. 
For $F_d = 0.1$, the value of $\Delta F$ is negligible compared to the work extracted, 
$-W$. Therefore the major contribution to $\Delta F - W$ comes from work extracted during 
the potential flip. The work done is $W \approx - 2 U_0 p(s)$, when the $S$ is a narrow 
region lying close to the maxima of the potential. Thus for low values of $s$, we have 
$\eta(s) = 2 U_0 p(s)/ \left<I(p(s))\right>$, which decreases as $p$ decreases. 
Since $p$ decreases with $s$, the downward trend of efficiency for small $s$ values is expected. 
At large values of $s$, an increase in $s$ leads to a decrease in $-W$ as the particle is 
more likely to be spotted in regions where the potential flip will lead to less work extraction. 
Given that $\left<I\right>$ is an increasing function of $s$ (when $p < 0.5$), 
it is expected that the efficiency decreases with $s$ for large $s$ values. 
The maximum value obtained for $\eta$ is $54.4 \%$, which is $33 \%$ more efficiency than for 
the $s$ value used in the previous section and also in the experimental study 
\cite{toyabe2010experimental}.

\begin{figure}
\includegraphics[scale=0.36]{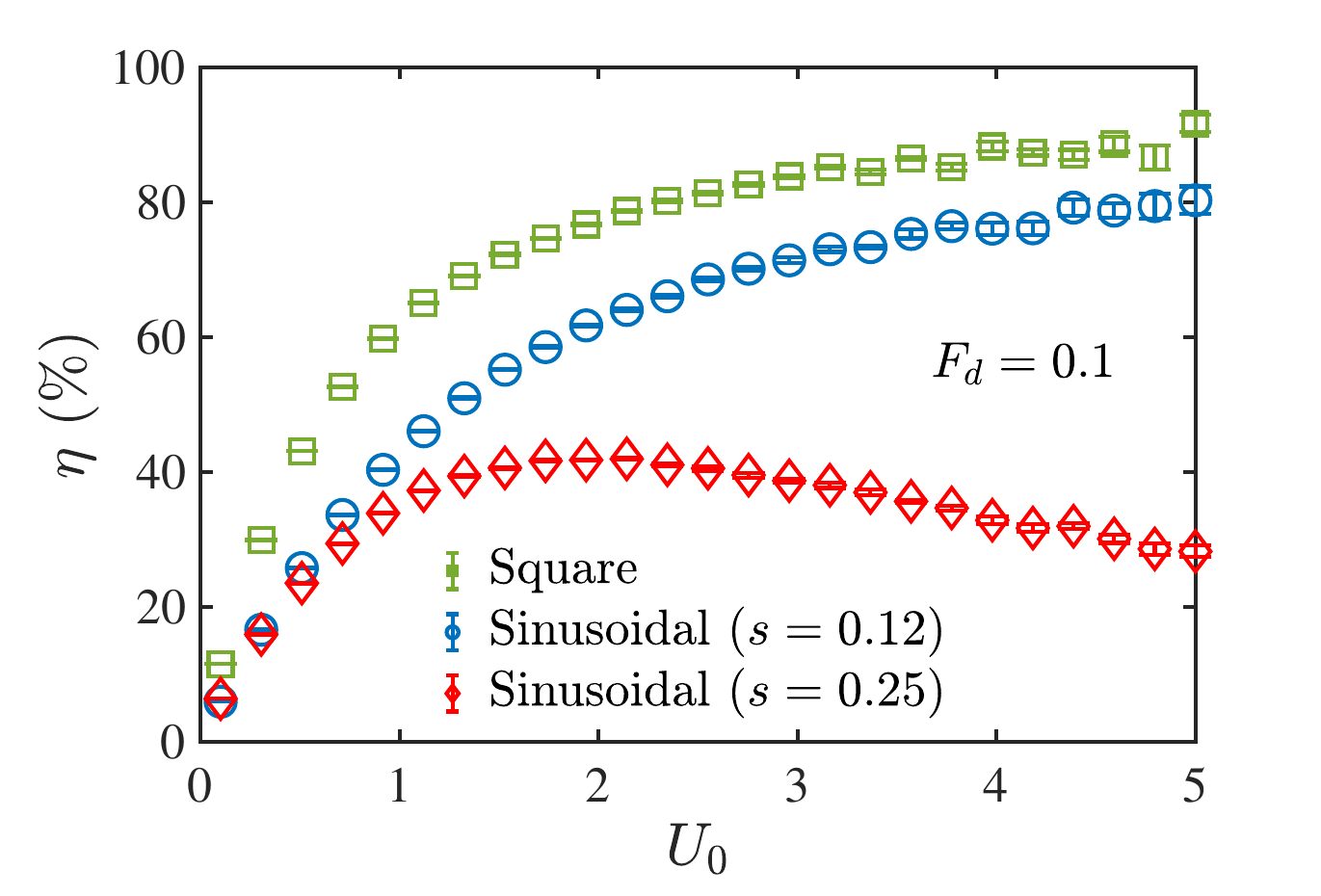}
\caption{\label{fig:optimization_A} Variation of efficiency with the amplitude for 
$s = 0.12$ (blue circles) and $s = 0.25$ (red diamonds). The maximum value of efficiency
for $s = 0.12$ is about $80 \%$. We have also shown the dependence of $\eta$ on amplitude
of the potential for the square shaped potential (green squares) with region $S'$ 
(shown in the inset of Fig. (\ref{fig:efficiency_square_potential}) in Sec. \ref{square}) 
as the searched region during measurement. The maximum efficiency for this case
is close to $90 \%$. The error bars in the figure are standard deviations.}
\end{figure}
\begin{figure}
\includegraphics[scale=0.36]{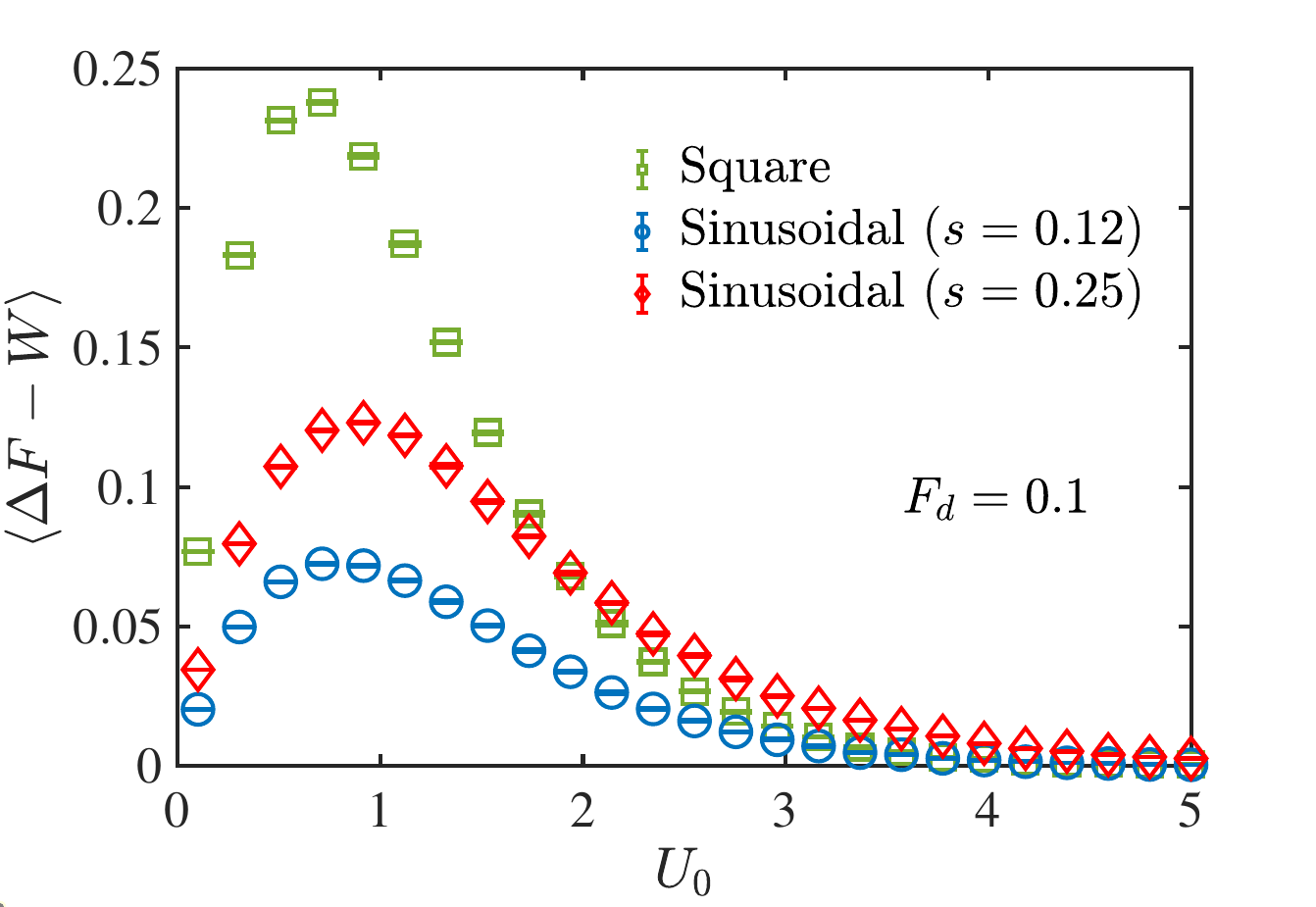}
\caption{\label{fig:power} Variation of total work extracted per cycle with amplitude of 
potential for $s = 0.12$ (blue circles) and for $s = 0.25$ (red diamonds). The functions have 
peaks close to $U_0 = 0.9$. Work extracted per cycle is more with $s = 0.25$. 
The work extracted per cycle as a function of $U_0$ for the case of square potential 
with $S'$ region as the search location during measurement (green squares) shows a 
much more prominent peak. The value location of the peak is marginally lower at $U_0 \approx 0.7$. The error bars in the figure are standard deviations.}
\end{figure}

Optimization of efficiency with respect to amplitude $U_0$ is done by keeping the value 
of $s = 0.12$, obtained above. As seen from Fig. (\ref{fig:optimization_A}), efficiency 
increases with amplitude and attains a maximum value of about $80 \%$ for amplitude 
$U_0 \gtrapprox 4.0$. The increase in efficiency with amplitude can be understood as 
follows: As $U_0$ increases the energy equivalent of available information, 
$k_B T \left<I\right>$ decreases because the probability to find the particle in 
region $S$ decreases. The average work extracted during the flip ($\left<-W\right>$), 
has a more complex dependence on $U_0$. The work done during individual flip of the 
potential will increase with $U_0$ on the average, but the occurrence of flips itself 
become exponentially less likely with increasing $U_0$. This leads to the behavior 
seen where $-W$ increases initially with $U_0$ but then falls to zero very fast as 
can be seen in Fig. (\ref{fig:power}). For low values of $U_0$ (compared to $k_BT$) 
the efficiency tends to zero since $-W$ is negligible and $k_B T \left<I\right>$ is 
finite. At large values of $U_0$, both the numerator and denominator in the expression 
for efficiency tend to zero, leading to a relatively flat curve. 
%In the numerical calculations it is seen that efficiency decreases with $U_0$ as $U_0$ 
%becomes larger than $k_B T$, at least for $s$ values that give us reliable statistics 
%(see Fig. (\ref{fig:optimization_A})). 

Even though there is a substantial increase in the efficiency attained by increasing the 
$U_0$ value, it comes with the cost of extremely low value for work extracted per cycle 
of the information engine. The low values of $p$ at large $U_0$ make the available 
information limited. The dependence of work extracted per cycle on the 
amplitude is show in Fig. (\ref{fig:power}). We have also shown in the same figure the 
behavior of work extracted per cycle with amplitude for $s = 0.25$. In fact, $s = 0.25$ 
gives better values for this quantity than the efficiency optimized parameter value of $s = 0.12$. 
In both cases, the maximum value of work per cycle is obtained for a potential 
amplitude of $U_0 \approx 0.9$, which is of the order of $k_B T$. The maximum value of work per 
cycle for optimized and non-optimized case are correspondingly $0.07$ and $0.12$ in 
relevant units.

\subsection{Feedback protocol studies}
As pointed out at the end of Sec. \ref{Results}, one of the reasons for 
the protocol used to be sub-optimal is that there is no feedback employed when 
the particle is \textit{not} spotted in $S$. By incorporating a feedback protocol 
for the negative result of the measurement outcome, we can improve the efficiency 
of the engine. One way to do this would be by raising the potential barrier height 
(deepening the well) when the particle is not spotted in $S$. This would increase 
the efficiency by mitigating the motion of particle in the direction of the drive. 
When the particle is not spotted in $S$, the chance that the particle is to the 
right of the potential minima (region $R$ in Fig. (\ref{meas})) is more than 
it is otherwise. So by raising the amplitude of the  potential, the possibility of 
the particle drifting in the direction of of the external drive is reduced,
thus helping in cutting the loss of free energy.

We have implemented this protocol with the parameters values kept same 
as in Sec. \ref{Results}. In the new protocol, in addition to
the potential flip carried out when the particle is seen in $S$, the 
amplitude of the potential is increased from $U_0$ to $1.1 U_0$, when it
is not. The increased value of amplitude is maintained till $t = 0.0490$ 
and the potential amplitude is reverted to its original value
at the end of the cycle at $t = \tau (0.05)$. Such a modification in 
the feedback protocol increased the efficiency marginally from around 
$41\%$ to about $43\%$. It is seen that the new protocol could marginally 
reduce the movement of the particle in the direction of drive. The gain 
in efficiency is not appreciable due to the fact that the changes in 
$\Delta F$ is not very crucial for the total work extracted, since the key 
contribution is coming from the $-W$ term.

\subsection {Square potential}
\label{square}
\begin{figure}
\includegraphics[scale=0.10]{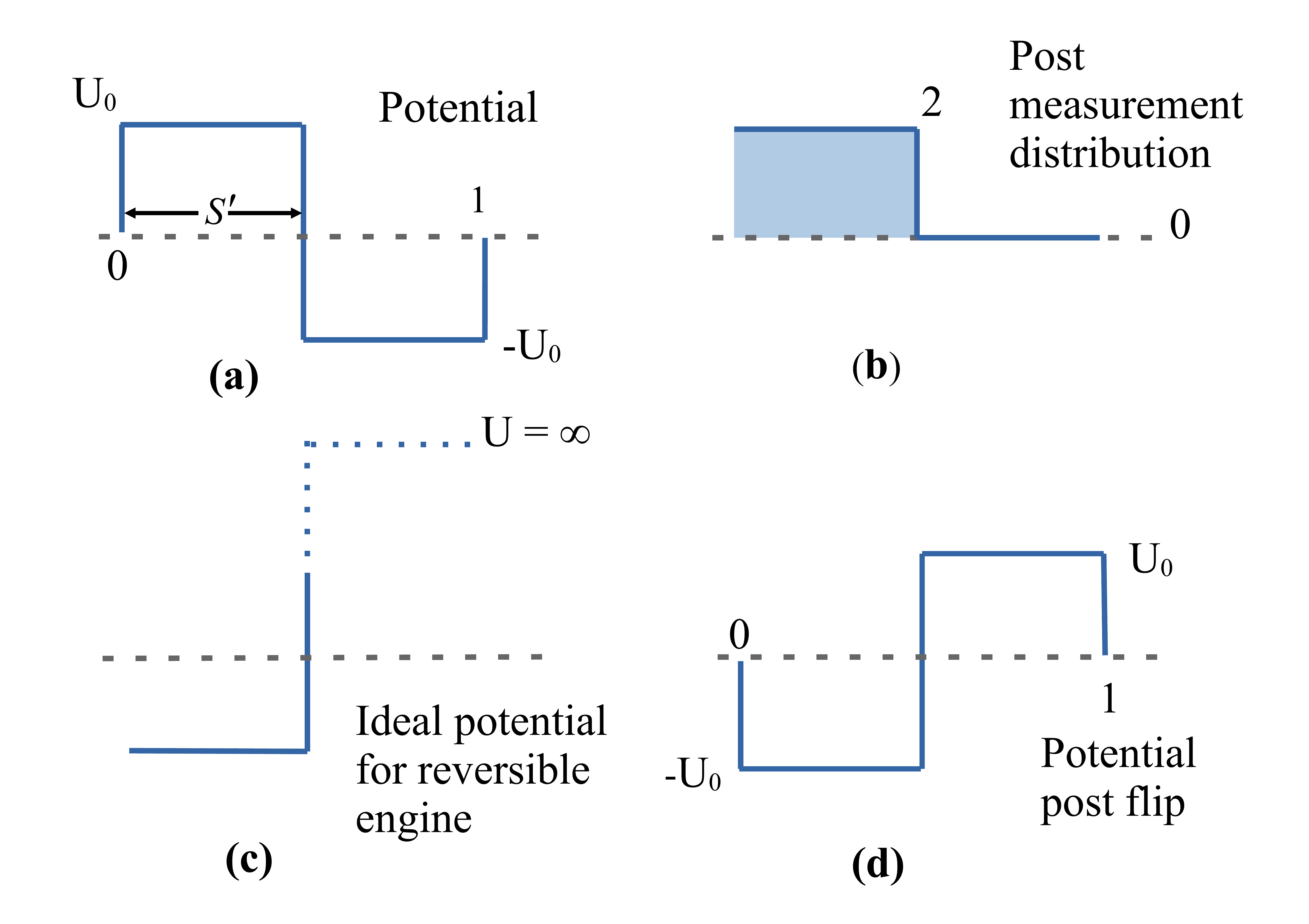}
\caption{\label{fig:whysquare}Feedback process with square periodic potential. 
\textbf{(a)} The initial shape of the potential energy function. $S'$ is the region where 
particle is probed for during the measurement. \textbf{(b)} Post measurement distribution of the 
particle for the case when particle is spotted in the region $S'$. \textbf{(c)} The 
potential that has its equilibrium distribution as the post measurement 
distribution for the case when the particle is spotted in $S'$. \textbf{(d)} The flipped potential 
at the end of the cycle.}
\end{figure}

We can extract maximum work using feedback by instantaneously changing 
the Hamiltonian of the system after the measurement such that the post 
measurement state of the system is identical to the equilibrium state of 
the new Hamiltonian \cite{esposito2011second,Parrondo2015}. 
Such a protocol avoids dissipation as the particle density does not have 
to relax to the new equilibrium distribution. If one now changes the potential
back to its original form quasi-statically, one can extract maximum work 
in the full cyclic process. In a scheme where the measurement amounts to finding the 
presence of the particle within a region and feedback protocol employs flipping 
of the potential, the choice of a square potential would come closest to 
achieving the above condition. This can be justified as follows: Consider a square
of period one given by,
\begin{eqnarray}
   U_s(x) &=& U_0 \;\;(0 < x \le 0.5)  \nonumber \\
   &=& -U_0 \;\;(0.5 < x \le 1),
\end{eqnarray}
with $U_s(x) = U_s(x+1)$ (see Fig. \ref{fig:whysquare}(a)). Consider
$S'$ to be the region between between $x = 0$ and $x = 0.5$ where the particle 
is searched for in the measurement phase. The post measurement density 
when the particle is spotted in $S'$ has a uniform value equals to $2$ in region $S'$ 
and zero outside of $S'$ (see Fig. \ref{fig:whysquare}(b)). To implement the 
reversible scheme, one would then have to switch the potential to an 
infinite barrier shape as shown in  Fig. \ref{fig:whysquare}(c) and then 
quasi-statically bring it back to the flipped potential shown in 
Fig. \ref{fig:whysquare}(d). Since we are restricting ourselves 
to potential flips, we by-pass the intermediate step. For large barrier 
height compared to $k_B T$, the loss incurred in this bye-passing of the 
intermediate procedure will be minimal because post measurement density
distribution will then be very close to the equilibrium distribution for
the flipped potential.  

\begin{figure}
\includegraphics[scale=0.4]{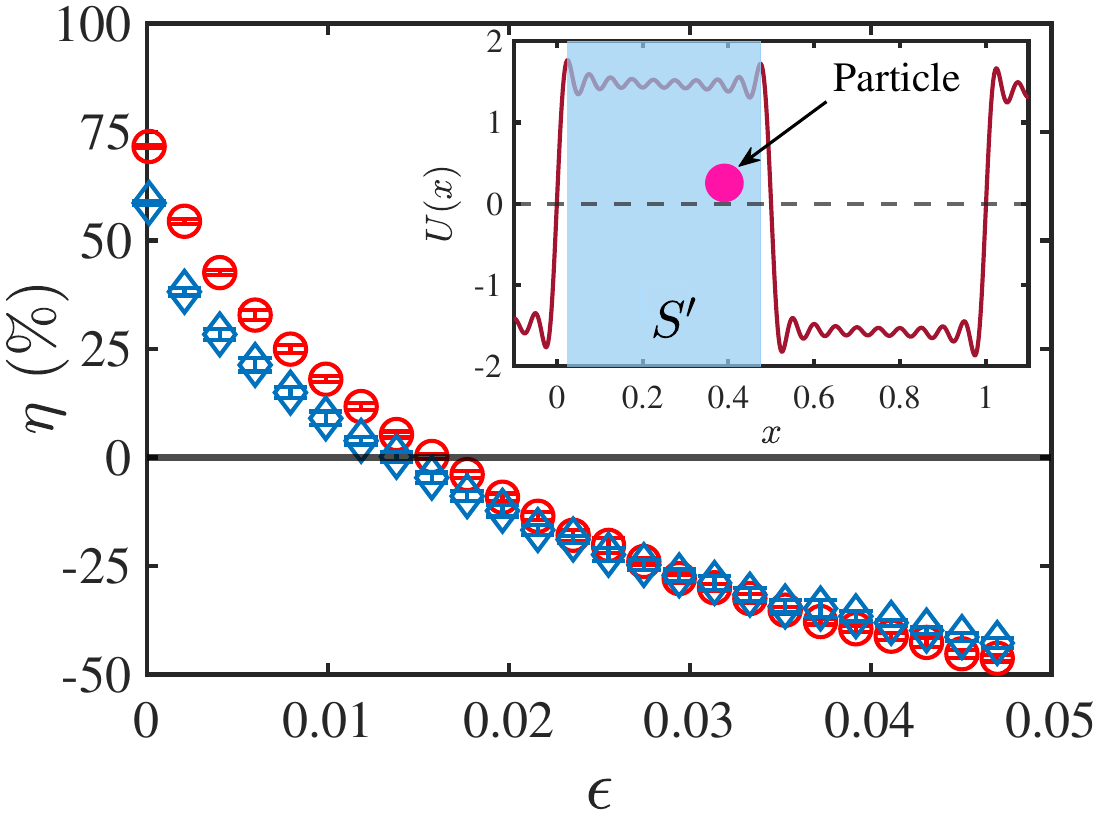}
\caption{\label{fig:efficiency_square_potential}Comparison of efficiency values 
with different values of $\epsilon$ in the case of square potential. The inset 
shows the newly defined region ($S'$) of particle's position measurement. 
$S'$ covers approximately the raised portion of the potential. 
The red circles correspond to efficiency values when the particle's position is 
measured in $S'$. The blue diamonds correspond to efficiency values when the 
particle's position is measured only in half of the raised portion of the 
potential. Such a region starts from the middle of the raised portion of the potential 
at $x = 0.24$ and ends at $x= 0.49$. It is seen that choosing the entire raised portion 
of the square potential as the measurement region ($S'$) gives better efficiency 
values comparing to measuring the particle's position only in half of the raised 
portion of the potential, for all values of $\epsilon$ where efficiency is positive. The error bars in the figure are standard deviations.}
\end{figure}

We have carried out the simulations by replacing the sinusoidal potential with an 
approximate form for square potential, obtained by keeping the first $10$ terms in 
the Fourier series expansion of a periodic square potential of amplitude $U_0$. That is,
$U(x) = \displaystyle\frac{4U_0}{\pi} \sum_{n=1 (odd)}^{19}\frac{1}{n}\sin{(2n\pi x)}$. 
The uniform external drive is kept the same as before. The resultant net potential 
is shown in the inset of Fig. (\ref{fig:efficiency_square_potential}). 
%\begin{figure}
%\includegraphics[scale=0.4]{potential_form_square.pdf}
%\caption{\label{fig:square_potential} Tilted square potential with period 
%$1$ and opposite phases experienced by particle. 
%Height and slope of the potential are $U_0$ = 1.5\ k_{B}T$ and 
%$F_d = 0.1\ k_{B}T$ respectively. The form of the square potential 
%is $\displaystyle\frac{4A}{\pi}\sum_{n=1 (odd)}^{19}\frac{1}{n}\sin{(2n\pi x)}$ 
%with a linear potential $-F_d x$.}
%\end{figure}
The region where particle is searched for during measurement is chosen to be 
the position values where the potential is approximately, $U_0$ ($S'$ in the figure). 
This region extends from $x = 0.03$ to $x = 0.47$ for the approximate square 
potential modelled above. We have also computed the efficiency for the case when
the region where particle is searched for is half of the raised portion of the 
potential. The variation of engine efficiency is shown in 
Fig. (\ref{fig:efficiency_square_potential}). 
For the smallest feedback delay, switching to square potential has increased the 
efficiency to almost $71\%$. Note that for the sinusoidal potential with same 
$U_0$ and optimal $s$ is only $54 \%$. Like in the case for sinusoidal potential, 
the contribution to the numerator of $\eta$ coming from $\Delta F$ is negligible 
compared to $-\left<W\right>$.

We have also studied the variation of the efficiency and power as a function of 
the amplitude of the square potential. The results are shown along with that for 
the sinusoidal potential (see Fig. (\ref{fig:optimization_A}) and 
Fig. (\ref{fig:power})). It is seen that the square potential offers better 
efficiency and work per cycle for almost the entire range of $U_0$. The maximal work
per cycle for the square potential is almost double of that obtained for the
sinusoidal potential with $s = 0.25$. The maximum efficiency for the square case 
is above $90 \%$ compared to the $80 \%$ for the sinusoidal one.

%\begin{figure}
%\includegraphics[scale=0.4]{velocity_square_potential.pdf}
%\caption{\label{fig:velocity_square_potential}  Variation of particle's 
%displacement rate (velocity) with feedback delay $\epsilon$ in a square potential. 
%On the average the particle is slightly moving towards the 
%right, though the information to energy conversion efficiency is high.}
%\end{figure}
%Furthermore, when the particle moves in the unflipped \textit{sinusoidal potential}, 
%the repulsive force on the particle due to the sinusoidal potential increases only 
%gradually before it reaches a maxima. Such a smoothly varying potential comparatively 
%increases the chances for the particle hopping to region $S$. However, in the case 
%of a square potential, such a hoping towards the left happens 
%with less probability. This is because the particle feels a sharp increase 
%in repulsive force with larger value when moves towards left 
%(See inset of Fig. (\ref{fig:gamma_square_potential})). This results in small 
%information gain and is comparable to the value of extracted work. Even though the 
%information available for conversion is very small, the rate of conversion of 
%information to energy (called efficacy) is larger in the case of square potential, 
%as discussed in the next section.

\section{Verifying generalized fluctuation theorems}
\label{vgift}
The generalized integral fluctuation theorem (GIFT) for processes which include an 
error-free feedback mechanism is given by \cite{ashida2014general}
\begin{equation}
\displaystyle \left<e^{(\Delta F - W)/k_{B}T - I + I_u}\right> = 1,
\label{gJeI}
\end{equation}
where the average is carried out over multiple trials, all starting off with the initial 
state of the system in equilibrium at temperature $T$. The quantity $I_u$ is the unavailable 
information associated with each measurement outcome. It is determined by running
the process backwards without feedback and finding the probability, $p_1$ ($p_2$) of finding 
the particle within region $S$ (outside of region $S$) if the particle was spotted in $S$ 
(not spotted in $S$) in the forward process. $I_u$ is given by $-\log(p_1)$ ($-\log(p_2)$) 
if the particle is spotted in $S$ (not spotted in $S$) in the forward process. 
The generalized form of Jarzynski equality (GJE) when feedback is present can be written as
\cite{sagawa2010generalized},
\begin{equation}
 \displaystyle \left<e^{(\Delta F - W)/k_{B}T} \right> = \gamma,
 \label{gJeGamma}
\end{equation}
where $\gamma = p_1 + p_2$. If the process is completely reversible, then $p_1 = p_2 = 1$ 
and we have the maximum value of $\gamma = 2$. The unavailable information in this case 
becomes zero. Thus $\gamma$ measures the efficacy of information to free-energy conversion 
and its value can vary between $0$ and $2$ for the present protocol, with two possible 
measurement outcomes used for the feedback process. One expects to regain the usual JE if 
there is no correlation between the outcome and the feedback, which is expected to
happen when the waiting time $\epsilon$ is comparable to the equilibration time. 
As the feedback procedure becomes more and more irreversible, one should even expect 
the $\gamma$ value to drop below $1$ due to the very low efficacy of the process.
\begin{figure}
\includegraphics[scale=0.4]{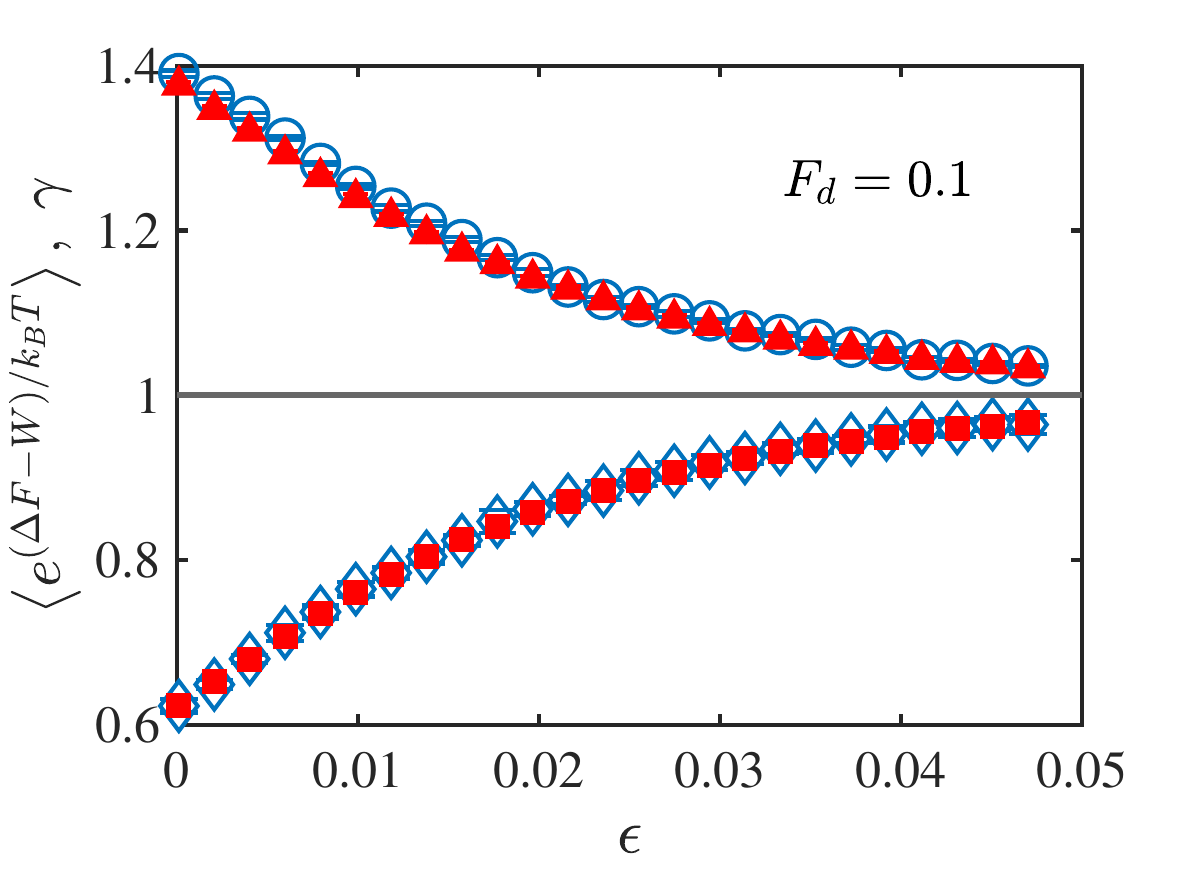}
\caption{\label{jarzy_sinu} Variation of left and right hand side of GJE 
with feedback delay,$\epsilon$, for a driving force value $F_d = 0.1$. 
The open circles give the LHS of Eq. (\ref{gJeGamma}) and the solid triangles 
the RHS. At large values of $\epsilon$ the feedback effect becomes minimal and 
conventional JE is approached. The open diamonds and solid squares too represent 
the same relation, but now for a feedback protocol that is designed to make the process more irreversible. The error bars in the figure are standard deviations.}
\end{figure}

We verify the GJE in the simulations by averaging over a large number of 
trials ($10^6$ runs for each value of $\epsilon$) that is warranted by the exponential 
average involved \cite{jarzynski2011equalities,liphardt2002equilibrium}. We have 
calculated $\gamma$ in our simulation as follows: The time reversed trajectories are 
obtained by running the process in time reversed manner with and without potential flip. 
The reverse cycle starts at $t = 0$ with the system in equilibrium and the potential is 
flipped at $t = \tau - \epsilon$ (for finding $p_1$) or not flipped (for finding $p_2$). 
At $t = \tau$, a measurement of the particle's position is done. If the potential is 
flipped in the reverse run, the probability for finding the particle in region $S$ is 
calculated, which gives $p_1$. For a run without potential flip, the probability for finding 
the particle outside $S$ is calculated, which is $p_2$. The sum of the two probabilities 
give us $\gamma$. The value of $p_1$ and $p_2$ are estimated by averaging over $10^6$ cycles
each and the error in their estimates is also determined. The average value of $p$ is 
determined using similar averages in the forward cycle.
\begin{figure}
\includegraphics[scale=0.4]{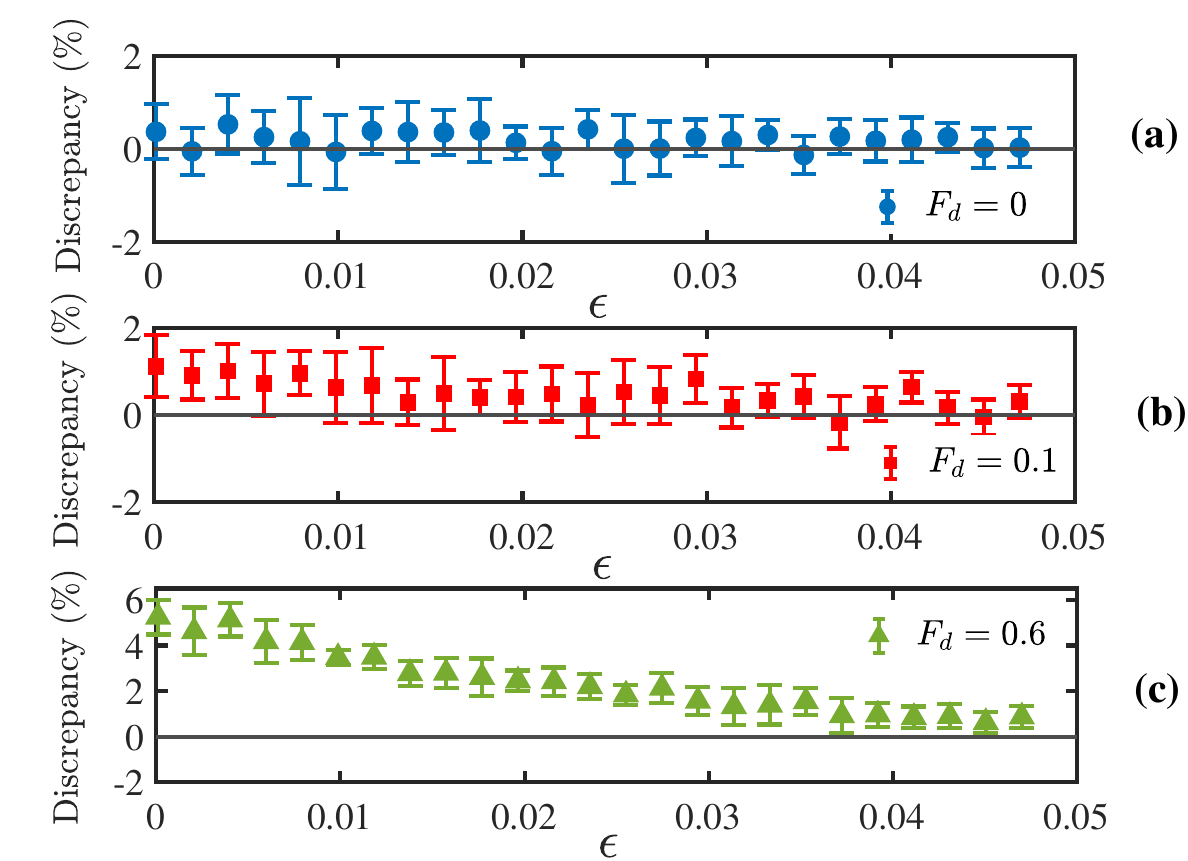}
\caption{\label{discrep} Discrepancy in percentage between the left and right 
hand side of GJE for \textbf{(a)} $F_d = 0$, \textbf{(b)} $F_d = 0.1$ and \textbf{(c)} $F_d = 0.6$. The discrepancy is 
calculated as $\left.\left[\left<e^{(\Delta F-W)/k_{B}T}\right>-\gamma\right] 
\middle/\left<e^{(\Delta F-W)/k_{B}T}\right>\right.$ in percentage. The error bars in the 
figure are standard deviations.}
\end{figure}

\begin{figure}
\includegraphics[scale=0.4]{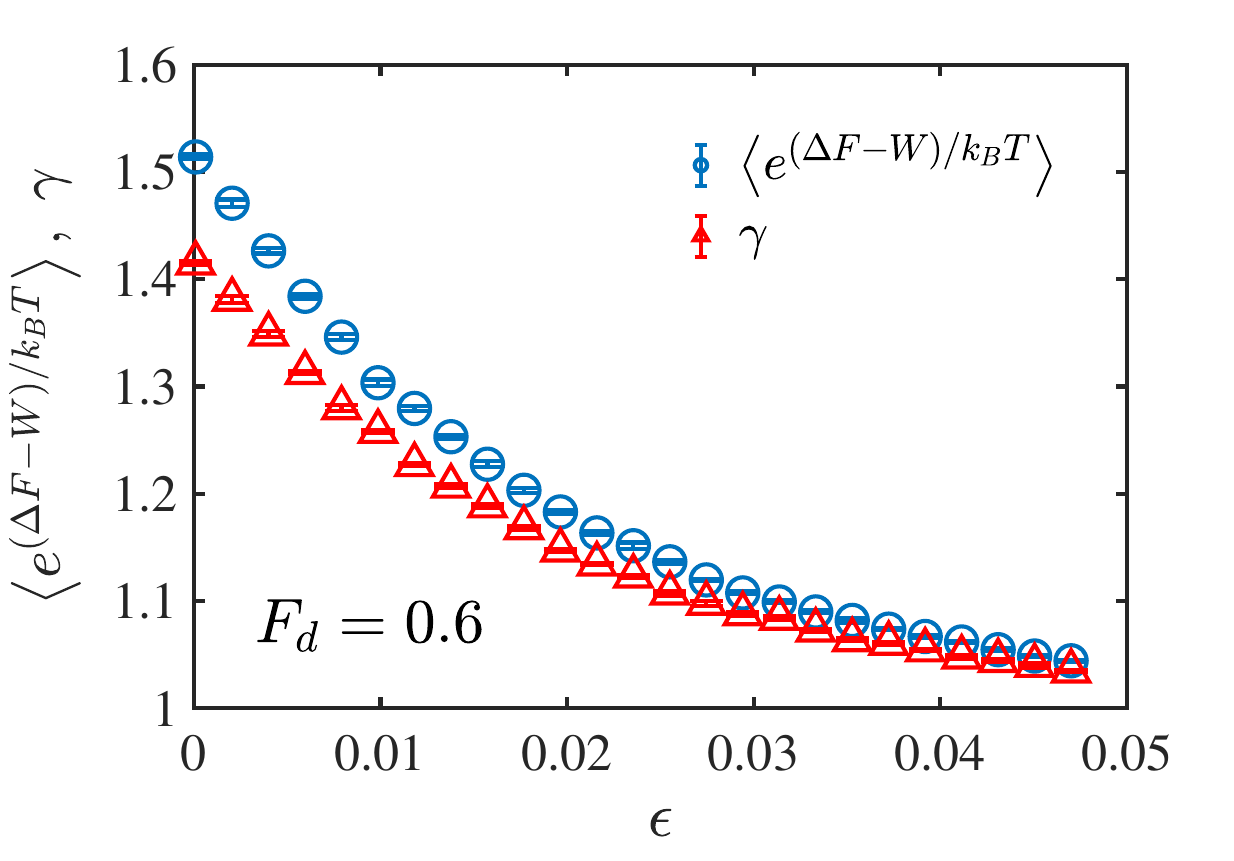}
\caption{\label{jarzy} Verification of GJE for larger 
driving force, $F_d = 0.6$. The discrepancy between the forward and 
reverse values are larger as compared to lower driving forces. The error bars 
in the figure are standard deviations.}
\end{figure}

\begin{figure}
\includegraphics[scale=0.4]{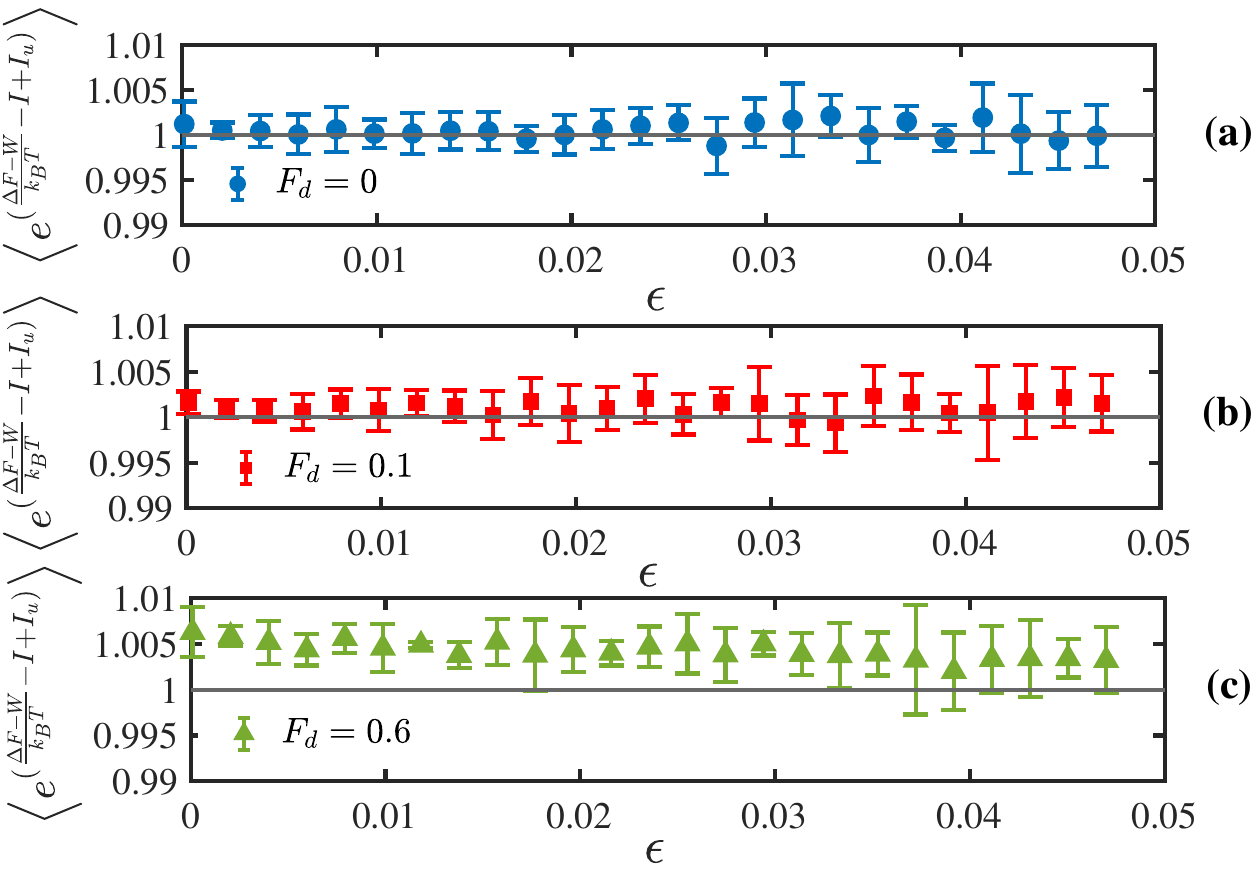}
\caption{\label{gift} The verification of generalized integral fluctuation 
theorem for the case with \textbf{(a)} $F_d = 0$, \textbf{(b)} $F_d = 0.1$ and \textbf{(c)} $F_d = 0.6$. There is a 
slight improvement of the validity of the relation for small drive. The values are 
visibly violating the relation for larger value of drive. GIFT is found to be valid 
for all values of $\epsilon$ studied with zero or small drive. The error bars in 
the figure are standard deviations.}
\end{figure}

Variation of left hand and right hand side of GJE as a function of $\epsilon$ for the 
sinusoidal potential case with $F_d = 0.1$ is shown in Fig. (\ref{jarzy_sinu}). 
As expected, the value of $\gamma$ starts above $1$ but below the maximum value of $2$ 
for small values of $\epsilon$ and is seen to tend towards $1$ for large values of $\epsilon$. 
The difference (in percentage) between the computed RHS and LHS of the equation is given in 
Fig. \ref{discrep} (b). The slight discrepancy that exists (less than $1 \%$), as seen 
from Fig. (\ref{discrep}) can be attributed to the fact that with the driving force on, 
the starting state is not a true equilibrium state. This mismatch is more apparent if 
the drive is stronger as can be seen from data in Fig. (\ref{jarzy}) which is for a 
driving force $F_d = 0.6$ and the corresponding discrepancy in shown in Fig. \ref{discrep} (c). 
Fig. \ref{discrep} (a) shows the difference between RHS and
LHS of GJE for $F_d = 0$. One can see that the difference has reduced and to within 
the statistical fluctuations the GJE is valid for all $\epsilon$ values.
The verification of GIFT carried out for $F_d = 0, F_d = 0.1$ and $F_d = 0.6$ are shown 
in Fig. (\ref{gift}) (a), (b) and (c) respectively. The relation is found to hold 
for all values of $\epsilon$ for the zero drive case. We have also verified the GJE for 
a feedback protocol that leads to values of $\gamma$ less than $1$. This was achieved 
by altering the feedback protocol such that the flip of the potential is carried out 
when the particles is {\it not} spotted in $S$ and the potential is left unaltered 
when the particle is spotted in $S$. The data given in Fig. (\ref{jarzy_sinu}) 
(bottom set) asserts the validity of the GJE for this case.

%We have seen earlier that replacing the sinusoidal potential with a square potential 
%gives a better efficiency. So one would expect that the process to be more reversible 
%and the 
%\begin{figure}
%\includegraphics[scale=0.4]{gamma_square.pdf}
%\caption{\label{fig:gamma_square_potential} Comparison of variation of left hand side 
%of generalized Jarzynski (efficacy) 
%with different values of feedback delay $\epsilon$ in the case of square potential.}
%\end{figure}
%For the same width of measurement $S$, we obtained a maximum efficacy around $1.58$ 
%for minimum feedback delay. The maximum value $2$ of the efficacy is obtained by measuring 
%the particle's position in the region $S'$ as described in 
%the previous section. Larger value of efficacy means that, even though the information 
%gain in square potential case is less (for a measurement over the region $S'$, $I =  0.198$), 
%the obtained information is effectively used for feedback controlling
%(Fig. (\ref{fig:gamma_square_potential})). Which means that 
%the rate of conversion of obtained information to extracted work is better in the 
%case of square potential.

\section{Conclusion}\label{conclusion}
We have carried out a Brownian dynamics simulation of a driven colloidal particle in 
one dimensional periodic potential with feedback control. An experimental study of a similar 
system using feedback control for converting the obtained information about the particle's 
position to free energy has been conducted before \cite{toyabe2010experimental}. The control 
over various model parameters as well as advantage of better averaging in simulations has 
allowed us to explore the model in great detail. Shorter waiting time for potential flip 
(in simulations, one can implement an almost instantaneous flip of the potential if the particle 
is spotted in $S$) as well as introduction of feedback process when the particle is not
observed during measurement allows one to reach an efficiency close to $43 \%$. Optimization
carried out by varying both the amplitude of the sinusoidal part of the potential $U_0$ as well 
as the width of the region $S$ in the model takes the efficiency value to $80 \%$. This is almost 
the double the value of $\eta$ before optimisation. But this comes at the cost of low value of 
resultant work extracted per cycle. For comparison, the experimentally obtained efficiency 
for a similar engine using colloidal particle in sinusoidal potential \cite{toyabe2010experimental} 
is $28\%$ and for an experimental implementation of Brownian motor working in a harmonic potential 
\cite{paneru2018optimal} is $35 \%$. 

The efficiency at maximum work per cycle that we have obtained is about $40 \%$ for the 
sinusoidal potential. It has to be kept in mind that the 
optimization has not been exhaustive and better combinations of power and efficiency could 
be possible. For comparison, the experimentally obtained efficiency at maximum power for
the experimental implementation of Brownian motor working in a harmonic potential referred to
above \cite{paneru2018optimal} is $19 \%$. It is not surprising that with the current model the 
conversion of all the available information to work is not possible. The fact that we are 
constraining the potential change after the measurement to the one arising from a flip 
does not allow one to tune the Hamiltonian post measurement to one where the post measurement 
distribution is the equilibrium distribution of the new Hamiltonian. This invariably leads to 
irreversibility in the process with the associated dissipation \cite{Parrondo2015}.

We have been able to work around the above limitation to an extent by using a square potential 
instead of the sinusoidal one. For large amplitude of the square potential, the flipping of the 
potential during the feedback process leads to a new Hamiltonian whose equilibrium distribution 
closely matches with the non-equilibrium distribution resulting from the measurement process. 
We find that both the work done per cycle as well as the efficiency has much better values for 
this choice. The efficiency goes above $90 \%$ for high values of amplitude and the efficiency 
at maximum work per cycle is $53 \%$. This suggests that using appropriate potential 
shape can lead to an appreciable change in the performance of this type of information engine 
based on a particle moving in a periodic potential.       

We have numerically verified GJE as well as GIFT for different values of feedback delay. 
The left and right hand sides of the GJE (see Eq. \ref{gJeGamma})  have been found 
independently using the forward and time reversed process respectively. We find that for 
zero drive force, fluctuation theorems are valid to within the simulation accuracy. 
The GJE for a similar system was verified experimentally \cite{toyabe2010experimental} 
and found to hold with $3 \%$ discrepancy. The error margin in the present simulation 
results are smaller with error bars down to less than $1\%$. We observe, like in the 
experiment, that for larger drives, the deviations in GJE is violated by a larger margin 
(around $5 \%$ for $F_d = 0.6$). This is expected since the condition of the starting state 
being an equilibrium one is then not met with \cite{Jarzynski1997,sagawa2010generalized}. 
Further, we have also verified GJE for the case when efficacy is a value less than $1$, 
implemented by changing the feedback process. It is found that for zero drive, 
the GIFT is valid for the waiting times studied. Though GIFT has been experimentally verified 
for and theoretically checked in information engine models based on a particle moving in a 
harmonic potential \cite{ashida2014general, paneru2018lossless}, this is the first 
time it is being verified for an error free information engine based on a particle 
moving in a periodic potential with feedback process based on potential flips. 

All the feedback studies we have carried out in this work are based on single 
cycle processes. The study can be extended by exploring the effects of correlation 
on efficiency of such an information engine by simulating multi-cycle realizations. 
There are preliminary indications that multi-cycle processes can lead to higher 
efficiencies. Experimental implementation of the information engine based on a 
colloidal particle moving in a square potential should be possible. 

\acknowledgements
The authors would like to acknowledge P. N. Deepak for fruitful discussions and
careful reading of the manuscript. TJ would like to acknowledge financial support 
under the DST-SERB grant CRG/2020/003646.

\bibliographystyle{apsrev4-2}
\bibliography{References}

\end{document}